\documentclass[a4paper]{article}

\usepackage{INTERSPEECH2020}
\usepackage[utf8]{inputenc}
\usepackage{array}
\usepackage{wrapfig}
\usepackage{multirow}
\usepackage{tabularx}
\usepackage{xcolor}
\usepackage{comment}
\usepackage{subcaption}
\usepackage{amsmath}
\usepackage{cite}
\usepackage{subcaption,graphicx}
\usepackage{caption}

\newcommand{\rulesep}{\unskip\ \vrule\ }

\title{Intra-class variation reduction of speaker representation \\in disentanglement framework}
\name{Yoohwan Kwon, Soo-Whan Chung and Hong-Goo Kang}
\address{
  Department of Electrical \& Electronic Engineering, Yonsei University, Seoul, South Korea}
\email{yhkwon@dsp.yonsei.ac.kr}

\begin{document}

\maketitle

\section{Abstract}
In this paper, we propose an effective training strategy to extract robust speaker representations from a speech signal.
One of the key challenges in speaker recognition tasks is to learn latent representations or embeddings containing solely speaker characteristic information in order to be robust in terms of intra-speaker variations.
By modifying the network architecture to generate both speaker-related and speaker-unrelated representations, we exploit a learning criterion which minimizes the mutual information between these disentangled embeddings.
We also introduce an identity change loss criterion which utilizes a reconstruction error to different utterances spoken by the same speaker.
Since the proposed criteria reduce the variation of speaker characteristics caused by changes in background environment or spoken content, the resulting embeddings of each speaker become more consistent.
The effectiveness of the proposed method is demonstrated through two tasks; disentanglement performance, and improvement of speaker recognition accuracy compared to the baseline model on a benchmark dataset, VoxCeleb1.
Ablation studies also show the impact of each criterion on overall performance.

\noindent\textbf{Index Terms}: speaker verification, disentanglement, mutual information

\section{Introduction}
Speaker recognition systems have been studied for many years due to their usefulness in various applications. 
Recently, the accuracy of speaker recognition has dramatically improved due to advances in deep learning and the availability of large-scale datasets for training.
The main objective of deep learning-based speaker recognition is to extract a high dimensional embedding vector such that it uniquely represents the characteristic of each speaker.
The d-vector~\cite{variani2014deep,chung2018voxceleb2} and x-vector~\cite{snyder2018x} are typical examples, where they are estimated via an identity classification task with an encoder style network.
The detailed extraction process differs with respect to the type of network structure and the criterion of the objective function such as softmax, triplet, and angular softmax~\cite{huang2018angular}.
However, given that the extracted embeddings also include speaker-unrelated information , there remains room for further improvement.

To overcome the aforementioned limitation inherent to the encoder style framework, a method for disentangling the embeddings with the use of relevant and irrelevant speaker information was proposed~\cite{Tai2020SEFALDRAS}.
The method consists of two encoders, a speaker purifying encoder and a dispersing encoder, as well as a decoder for reconstruction.
While the speaker purifying encoder is trained by the original speaker classification scheme, the dispersing encoder is trained by an adversarial training scheme designed to fool it from correctly classifying the speaker identity.
Later, two encoded features are concatenated, following which they are fed to the decoder, which utilizes a reconstruction loss to the original input so that all information is embedded within the representative features.
In other words, they decompose the entirety of the speech information into speaker identity-related and -unrelated information.
Although the speaker and non-speaker embeddings are learned effectively using the adversarial classifier, the method does not directly address the task of dispersing both embeddings simultaneously in disentanglement.
There is an opportunity to improve the disentanglement performance by adopting a method which considers the relation of embeddings simultaneously.

In this paper, we propose a method to effectively disentangle speaker identity-related and identity-unrelated information using various types of criteria.
We first introduce a criterion for minimizing mutual information between speaker-related and -unrelated representations, which is beneficial due to that it directly considers the relation between those features.
We also propose a novel identity change criterion which measures the difference between the input and generated mel-spectrums.
The reconstructed mel-spectrum used for the identity change loss is generated via a speaker embedding from one utterance and a residual embedding from the other utterance possessing the same speaker identity. 
Since the criterion enforces speaker embeddings to be similar to a different set of utterances, it reduces intra-variation within each speaker's cluster.
The main contributions of this paper are as follows: (1) we propose an effective method for disentanglig identity-related and identity-unrelated information using a mutual information criterion through an auto-encoder framework; (2) we introduce a speaker identity change loss criterion to further enhance the performance of speaker embeddings; (3) we use this framework to improve speaker verification performance on benchmark datasets.

The remainder of the paper is organized as follows.
Section 3 presents a brief overview of related works on speaker embedding and disentanglement.
In Section 4, we present the details of the proposed method such as network architectures and loss functions.
Experimental results are presented in Section 5, and the conclusion follows in Section 6.

\section{Related works}

\subsection{Speaker embedding strategy}
Speaker embedding vectors are high level representations (typically obtained via deep neural networks) that aim to compactly represent a speaker's identity. They are very important for many applications such as speaker recognition and diarization.
There are various speaker embedding methods that differ in terms of the type of network architecture, feature aggregation, and training criteria.
Deep learning architectures such as DNN-~\cite{variani2014deep,heigold2016end,snyder2016deep}, CNN-~\cite{nagrani2017voxceleb,chung2018voxceleb2,li2017deep, hajibabaei2018unified,jungimproving}, or LSTM-based ones~\cite{wan2018generalized} first extract the frame-level features from a variable length of utterances.
Then, a pooling method~\cite{snyder2017deep, cai2018exploring,cai2018analysis, xie2019utterance} is used to aggregate the frame-level features to a fixed length of utterance-level.
In terms of the objective function, they are trained by performing a classification task with a criterion of softmax, angular softmax or a metric learning task using a contrastive loss~\cite{nagrani2017voxceleb, chung2018voxceleb2}, triplet loss~\cite{li2017deep} and etc~\cite{chung2020defence, kye2020meta}.
Nevertheless, there is still room for improvement if we introduce the concept of target-unrelated information to the extracted embedding features.

\subsection{Disentangled feature learning}
Disentanglement is a learning technique that represents the input signal's characteristics through multiple separated dimensions or embeddings.
Therefore, it is beneficial for obtaining representations that contain certain attributes or for extracting discriminative features. 
Adversarial training~\cite{ganin2016domain,zhou2019training, meng2019adversarial, peng2019domain,bhattacharya2019generative} and reconstruction based training~\cite{zhang2019non, chou2018multi,liu2018exploring,eom2019learning,gonzalez2018image} are widely used to obtain disentangled representations.

Tai at el.~\cite{Tai2020SEFALDRAS} proposed a disentanglement method for speaker recognition that is the baseline for our work.
By constructing an identity-related and an identity-unrelated encoder, they trained each encoder to represent only speaker-related and -unrelated information using speaker identification loss and adversarial training loss.
They also adopted an auto-encoder framework to maintain all input speech information within output embeddings. 
The information contained in the output embeddings is preserved using spectral reconstruction approaches.

\subsection{Mutual Information Neural Estimator}
Mutual information (MI) based feature learning methods have been popular for a long time, but they are often difficult to apply for deep learning-based approaches because it is not easy to calculate the MI for high dimensional continuous variables.
Recently, a mutual information neural estimator (MINE)~\cite{belghazi2018mine} was proposed to estimate mutual information with a neural network architecture.

By definition, the MI is equivalent to the Kullback-Leibler (KL) divergence of a joint distribution, $P_{X,Y}$, and the product of marginals, $P_{X \otimes Y}$.
According to the Donsker-Varadhan representation~\cite{donsker1983asymptotic}, the lower bound of mutual information can be represented by:
\begin{equation}
  I(X,Y) \geq \sup_T\mathbb{E}_{P_{X,Y}}[T_\theta]-log(\mathbb{E}_{P_{X \otimes Y}}[e^{T_\theta}]).
 \label{mine}
\end{equation}
The $T$ function is trained by a neural network with the parameter $\theta$, for which the output can be considered to be an approximated value of mutual information between $X$ and $Y$.
It has been widely used in recent works on feature learning~\cite{hjelm2018learning, ravanelli2018learning, sanchez2019learning}.
\begin{figure*}[t]
\centering
\begin{subfigure}{.19\textwidth}
  \centering
  \includegraphics[width=\linewidth]{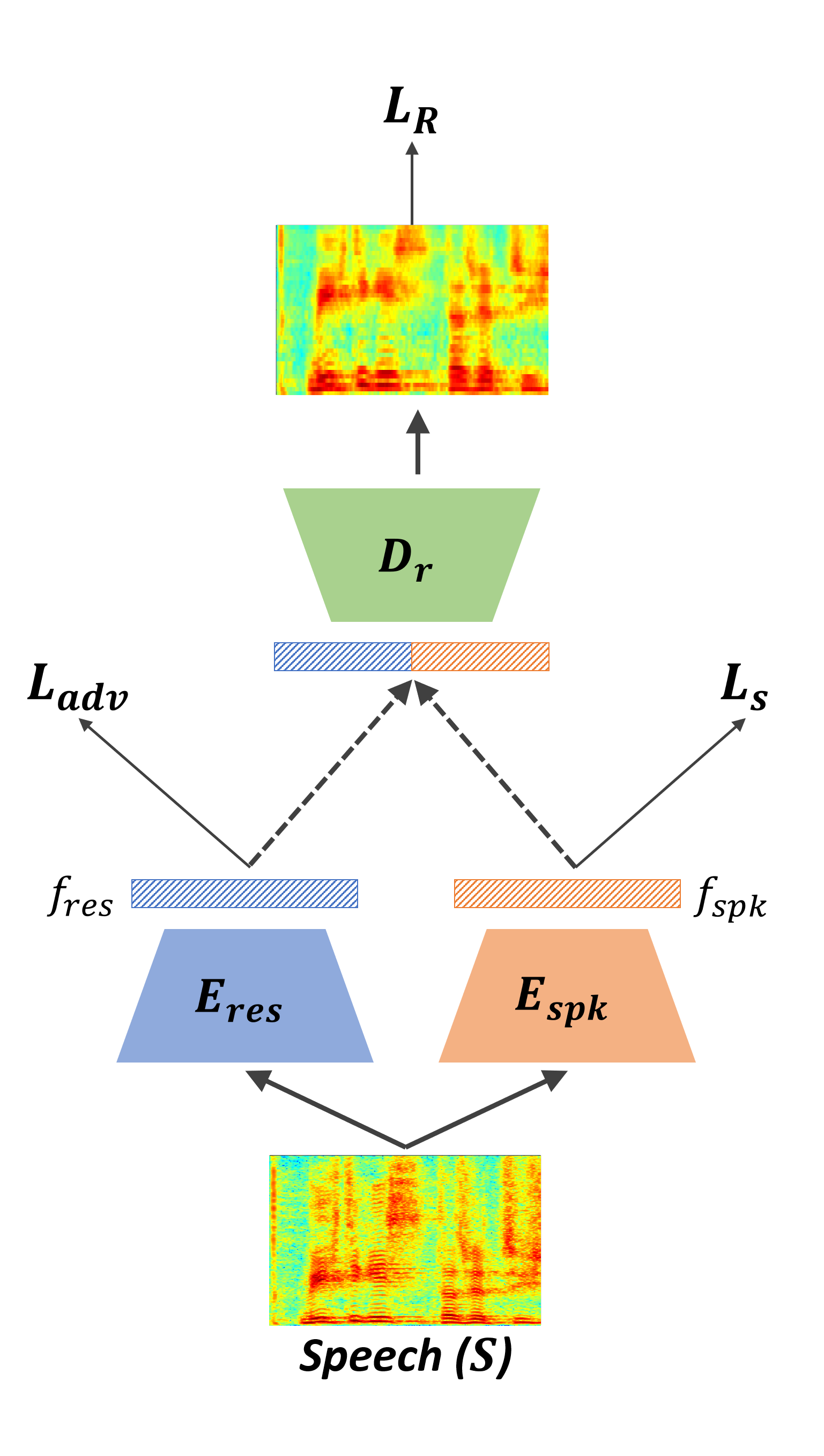}  
  \caption{Baseline loss}
  \label{baseloss}
\end{subfigure}
\rulesep
\begin{subfigure}{.418\textwidth}
  \centering
  \includegraphics[width=\linewidth]{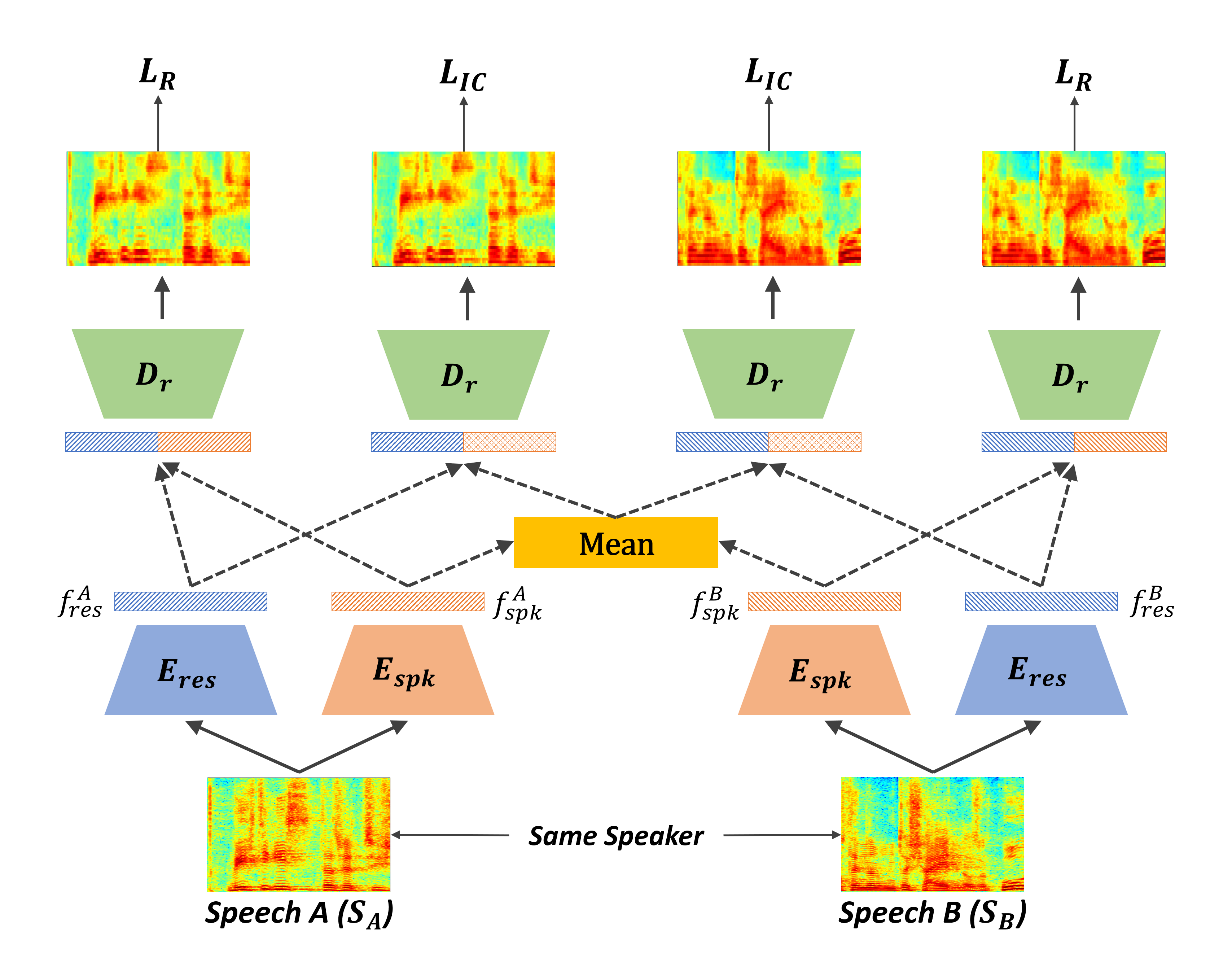}  
  \caption{Identity change loss}
  \label{idloss}
\end{subfigure}
\rulesep
\begin{subfigure}{.324\textwidth}
  \centering
  \includegraphics[width=\linewidth]{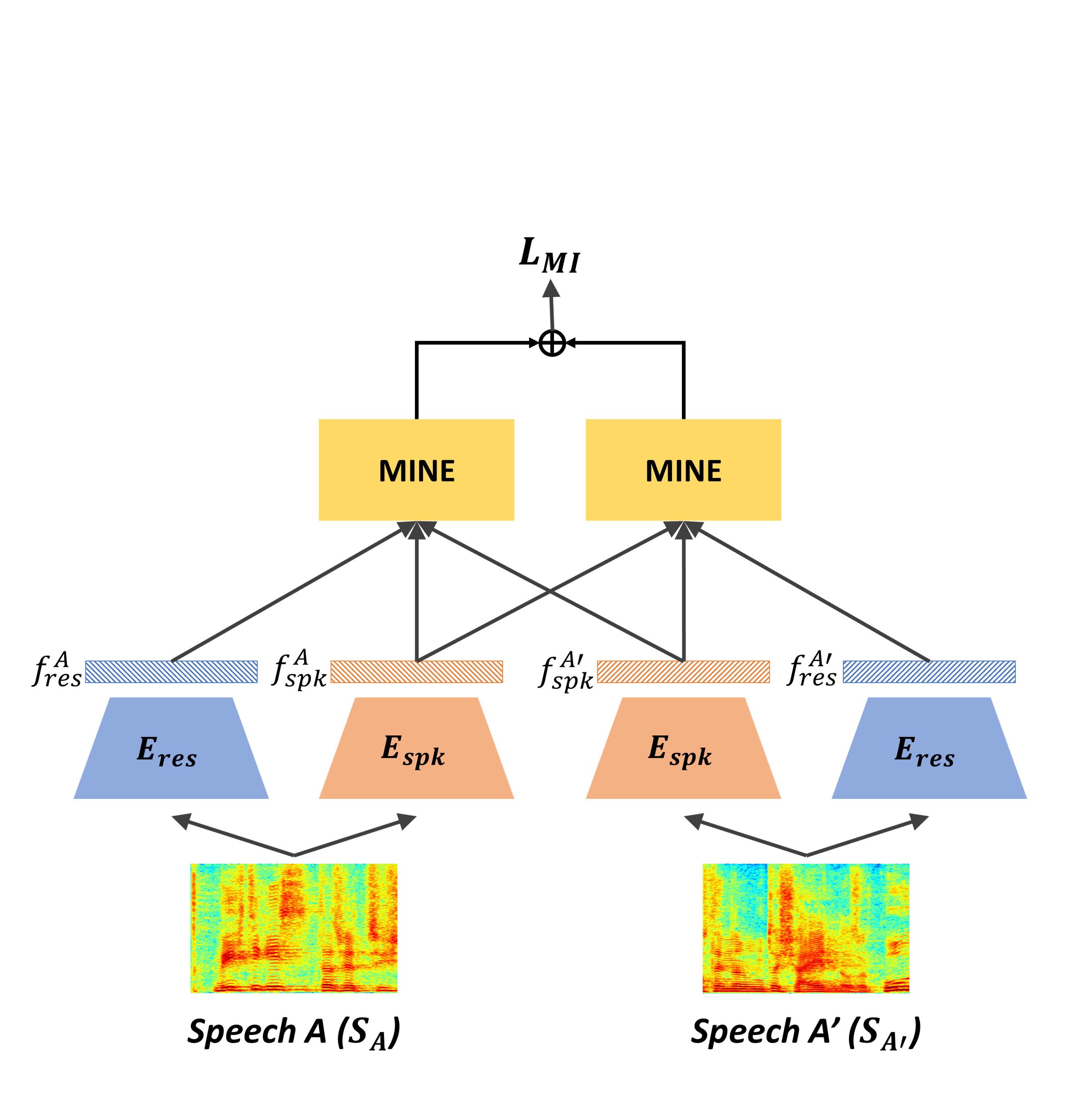}  
  \caption{Mutual information loss}
  \label{mineloss}
\end{subfigure}
\caption{Overview of proposed training criteria.
{\normalfont(a)} Training criteria based on~\cite{Tai2020SEFALDRAS}: speaker loss, disentanglement loss and reconstruction loss. 
{\normalfont(b)} Identity change loss: switch the speaker embedding to mean of those.
{\normalfont(c)} Mutual information loss: estimate the mutual information from speaker and residual embeddings by MINE} 
\vspace{-10pt}
\label{fig:overall}
\end{figure*}

\section{Proposed Method}
The main goal of the proposed algorithm is to extract a high-level latent embedding that contains only speaker-related information.
To achieve this goal, we propose a disentanglement method to decouple speaker information from an input signal such that the embedding represents the speaker's identity being robust to the variation of linguistic information.

\subsection{Overview of the proposed algorithm}
Figure~\ref{fig:overall} illustrates the proposed training strategies in our disentanglement method.
Our network consists of three modules: a speaker encoder $E_{spk}$, a residual encoder $E_{res}$, and a decoder $D_r$.
$f_{spk}$ and $f_{res}$ are respectively the output features of encoders $E_{spk}$ and $E_{res}$.
Our method reconstructs mel-scaled spectrum instead of the magnitude spectrum so that it efficiently disentangles embeddings without speaker information loss.

The network is trained in various learning criteria used in the baseline model, depicted in Figure~\ref{baseloss} with auxiliary loss which minimizes intra-variance of clusters; speaker loss, disentanglement loss, reconstruction loss, and our novel criterion -- {\em identity change} loss.
Also, we modify disentanglement loss, which uses the adversarial classifier on the residual embedding in the baseline method, into the mutual information between $f_{spk}$ and $f_{res}$.
Details of each criterion are described in the Section~\ref{subsec:training_objective}.

\subsection{Training Objective}
\label{subsec:training_objective}
In this section, we demonstrate the details of the proposed method with objective functions for training; speaker loss $L_S$, disentanglement loss $L_{MI}$, reconstruction loss $L_R$ and identity change loss $L_{IC}$ .
The total objective function of the proposed method consists of four loss functions:
\begin{equation}
  \begin{split}
      L_{total}=&\lambda_1L_S+\lambda_2L_{MI}+\lambda_3L_R+\lambda_4L_{IC}. 
  \end{split}
  \label{total}
\end{equation}
The hyper-parameters are set based on experimental results, $[\lambda_1, ... ,\lambda_4] = [1, 0.1, 0.1, 0.1]$.

\vspace{2pt}
\noindent\textbf{Speaker loss.}
The objective of the speaker loss is embedding speaker representation $f_{spk}$ into the latent space using the encoder $E_{spk}$ as done in~\cite{ nagrani2017voxceleb, huang2018angular, li2017deep, wan2018generalized}.
Following the baseline model, the speaker encoder is trained in a speaker label classification task using a cross-entropy criterion.
The loss function is denoted as:
\begin{equation}
  L_{S}=-\sum_{i=1}^C t_i log(softmax(f_{spk})_i),
\end{equation}
where $C$ is the number of speakers and $t$ is the label index.

\vspace{2pt}
\noindent\textbf{Disentanglement loss.}
In the disentanglement mechanism, the residual embedding $f_{res}$ contains information which is not included in the speaker vector $f_{spk}$.
The baseline method adopts the adversarial classification to embed residual of speaker characteristics.
The adversarial classification shares network parameters used in speaker loss whereas its objective is to eliminates the speaker information by fooling the classifier.
The residual encoder $E_{res}$ is trained not to estimate any speaker label by using uniform distribution, and its definition is as follows:
\begin{equation}
    L_{adv}= {1 \over C} \sum_{j=1}^{C}log(\emph{softmax}(f_{res})_j),
\end{equation}
where $C$ is the number of classes.

In our strategy, we attempt disentanglement using mutual information between $f_{spk}$ and $f_{res}$ instead of adversarial learning.
Since the {\em genuine} disentanglement is achieved in dispersing residual information but not in embedding features separately, we consider both $f_{spk}$ and $f_{res}$ in terms of disentanglement criterion.
Here, we adopt the MINE method, which handles correspondence between the three embeddings using deep learning approaches.
In~\cite{ravanelli2018learning}, MINE controls the information differences between speakers; minimizing in the same speaker and maximizing in different speakers.
MINE, in our paper, maximizes the discrepancy between disentangled features ($f_{spk}$, $f_{res}$), and minimizes between speaker representations extracted from different segments of the same speech signals as shown in Figure~\ref{mineloss}.
The criterion is designed as Equation~\ref{eq:mi_loss}.
\begin{equation}
    \label{eq:mi_loss}
    \begin{split}
        L_{MI} = \mathbb{E}[T_\theta(f_{spk}^{A},f_{spk}^{A'})] -log\Big(\mathbb{E}\Big[e^{T_\theta(f_{spk}^{A},f_{res}^{A})}\Big]\Big) \\
                + \mathbb{E}[T_\theta(f_{spk}^{A'},f_{spk}^{A})] -log\Big(\mathbb{E}\Big[e^{T_\theta(f_{spk}^{A'},f_{res}^{A'})}\Big]\Big),   
    \end{split}
\end{equation}
where $f_{spk}^A$ and $f_{spk}^{A'}$ represent identical speaker extracted from the same speech signal with different offsets, and $f_{res}^A$ and $f_{res}^{A'}$ are their residual embeddings.
It holds the common information between speaker embeddings and disperses residuals to speaker embeddings on the other embedding.

\vspace{2pt}
\noindent\textbf{Reconstruction loss.}
The disentangled embeddings, $f_{spk}$ and $f_{res}$ preserve the spectral information in the input spectrum when they are combined. The decoder $D_r(f_{spk},f_{res})$ is trained to generate a reconstructed spectrum using a concatenated embedding input.
The reconstruction loss $L_R$ is defined by measuring the distance between input and the reconstructed spectrum using an MSE criterion as follows:
\begin{equation}
  L_{R}=||D_r(f_{spk},f_{res}) - S_{mel} ||^2,
\end{equation}
where $S_{mel}$ is a mel-spectrum of the input speech signal $S$.
Reconstructing the mel-spectrum instead of a magnitude spectrum can reduce the burden of the decoder during the spectrum generation process, while it still enables the generation of embeddings containing all information of input.

\vspace{2pt}
\noindent\textbf{Identity change~(IC) loss.}
Intra-class variance inevitable in each speaker cluster is caused by the variation of linguistic information, recording environments, and speakers' emotional or health state. 
To further improve speaker recognition performance by minimizing intra-class variances in speaker clusters, we propose identity change loss.
Instead of minimizing intra-class variance directly, we use a reconstruction loss criterion that measures spectral distance between the reference and reconstructed one.
Since the reconstructed mel-spectrum is generated by substituting the identity embedding with the one extracted from different utterances spoken by the same speaker, we may obtain perfect reconstruction only when the substitute embedding has the same distribution as the original identity.
The identity change loss is described in Equation~\ref{eq:ic_loss}.
\begin{equation}
 \begin{split}
     L_{IC}=&\|{\hat{S}_{A}-S_A}\|^2 +\|{\hat{S}_{B}-S_B}\|^2,\\
     \hat{S}_{A}=&D_r\bigg(\frac{f_{spk}^A+f_{spk}^{B}}{2}, f_{res}^A\bigg),\\
     \hat{S}_{B}=&D_r\bigg(\frac{f_{spk}^A+f_{spk}^{B}}{2}, f_{res}^B\bigg),
 \end{split}
 \label{eq:ic_loss}
\end{equation}
where $S_A$ and $S_B$ are the mel-spectrum of speech signals $A$, $B$ spoken by the same speaker, and $\hat{S}_A$ and $\hat{S}_{B}$ are the reconstructed mel-spectrum using substituted identities.
In the proposed method, $f_{spk}^A$ and $f_{spk}^B$ are substituted with the mean of two identities depicted in Figure~\ref{idloss}; it guides the direction where speaker embeddings to be gathered to minimize intra-class variance.

\begin{table}[t]
\centering
\caption{Verification results on VoxCeleb1 test set. S, C and AM are Softmax, Contrastive and Angular margin loss, respectively.}
\begin{tabular}{c | c c c c} 
 \toprule
                                              &\bf Model & \bf Criterion &\bf EER \\
 \midrule\midrule
 Chung \emph{et al.}~\cite{chung2018voxceleb2} & Encoder& S + C & 5.04\% \\
 Xie \emph{et al.}~\cite{xie2019utterance} & Encoder & S & 5.02\% \\
 \midrule
 Tai \emph{et al.}~\cite{Tai2020SEFALDRAS} & Enc(2)+Dec& S & 3.83\% \\

 \midrule
 \multirow{2}{*}{\bf Proposed}  & Enc(2)+Dec & S & \textbf {3.18\%}  \\
                               & Enc(2)+Dec & AM & \textbf {2.54\%}  \\
 \bottomrule
\end{tabular}
\vspace{-10pt}
\label{table:1}
\end{table}
\section{Experiments}

\subsection{Dataset configuration}
We train our model on VoxCeleb2~\cite{chung2018voxceleb2}, which is a large-scale audio-visual dataset containing over 1 million utterances for 5,994 celebrities, extracted from YouTube videos.
We evaluate our model on VoxCeleb1~\cite{nagrani2017voxceleb} test set which consists of 677 clips spoken by 40 speakers.
Clips are segmented into 3 seconds with a random offset from each utterance for training.
They are sliced in every 10ms with 25ms window length and transformed into log-magnitude spectrum with the FFT size of 512; thus, the dimension of input speech features is $300 \times 257$.
For reconstruction, we prepare mel-spectrogram in logarithm scale using 64 mel-filterbanks as outputs.

\subsection{Implementation details}
The structures of the speaker encoder and the residual encoder are designed based on ResNet34 with small changes into the pooling strategy.
Both encoders use a time average pooling (TAP) method to embed variable length input features into a fixed dimension of utterance level.
The decoder consists of 3 fully-connected layers and 9 transposed convolutional layers referenced by~\cite{radford2015unsupervised}.
In the training phase, the batch size of the input is set to 32 and the model is trained with the Adam optimizer~\cite{kingma2014adam}. 
The learning rate is set to 1e-3 and reduced by half every 10 epochs until convergence.

\begin{table}[t]
\centering
\caption{Ablation study of the proposed method}
\begin{tabular}{ c | c c c c c | c } 
     \toprule
     & $L_s$ & $L_r$ & $L_{adv}$ & $L_{mi}$ & $L_{ic}$ & EER (\%) \\
     \midrule
     Baseline & \checkmark & \checkmark & \checkmark & - & - & 3.83\% \\
     \midrule
     \multirow{4}{*}{\bf Proposed}
              & \checkmark & \checkmark & \checkmark & \checkmark &      -     & 3.71\% \\
              & \checkmark & \checkmark &      -     & \checkmark &      -     & 3.81\% \\
              & \checkmark & \checkmark & \checkmark &      -     & \checkmark & 3.59\% \\
              & \checkmark & \checkmark &      -     & \checkmark & \checkmark & \bf 3.18\% \\
     \bottomrule
\end{tabular}
\label{table:2}
\vspace{-10pt}
\end{table}

\subsection{Training strategy}
\noindent \textbf{Phase I. Disentanglement training.}
In phase I, the network is pre-trained using speaker loss, disentanglement loss and reconstruction loss, similar to the baseline strategy.
According to each experimental setup, either adversarial loss or mutual information loss is used.

\vspace{2pt}
\noindent \textbf{Phase II. Identity change training.}
During phase II, we consider an efficient training strategy for identity change loss.
Its motivation is based on dispersing information by setting one embedding as an anchor and stable adaptation of the other side embedding.
The detailed process is shown below and the stages are processed recursively:

\begin{enumerate}
\item {\em Intra-class minimization} -- The identity is replaced by the mean of two identities to generate mel-spectrogram, and its reconstruction error $L_{IC}$ is minimized through backpropagation on the decoder and residual encoder.
\item {\em  Adaptation} -- The original identity is ingested on the decoder and the parameters of the decoder and the speaker encoder are updated to minimize reconstruction error $L_{R}$.
\end{enumerate}

\subsection{Experimental results}
We compare the performance of our models to that of conventional models and analyze the impact of each loss function on overall performance with an ablation study under the same settings.
All models for comparison are re-implemented by ours.
Table~\ref{table:1} shows the equal error rate (EER) obtained by the VoxCeleb1~\cite{nagrani2017voxceleb} testset, where we compare our models with the encoder model~\cite{xie2019utterance} and the disentanglement model~\cite{Tai2020SEFALDRAS}.
With the standard softmax loss and TAP aggregation, our model outperforms previous models based on the ResNet encoder by 36.6\%  and the disentanglement model using an adversarial method~\cite{Tai2020SEFALDRAS} by 16.9\%.
These results demonstrate that the represented embeddings of the proposed disentanglement approach are more informative than those of the baseline. 
The proposed method trained with angular margin softmax provided our best results among the experiments.

\vspace{2pt}
\noindent\textbf{Ablation study.}
Table \ref{table:2} shows equal error rates (EERs) obtained by ablation studies, which indicates the effectiveness of loss functions used in the proposed model.
First, we trained the model using the mutual information criterion with and without the adversarial criterion.
The results confirm that minimizing the mutual information between speaker and residual embeddings is effective to disentangle speaker information.
Unlike adversarial training, which is applied to the encoders independently, mutual information is calculated between speaker and residual embedding simultaneously, resulting in more powerful disentanglement performance.
Among these experiments, the case absent adversarial criterion performs better, with an EER~3.81\%.
Then, the other experiments are conducted in order to investigate the effect of identity change loss.
The results prove that identity change loss improves the performance of speaker embedding, and it shows the best result when it is trained using the mutual information and identity change loss criterion together, giving an EER~3.18\%.

\begin{figure}[t]
\begin{subfigure}{.23\textwidth}
  \centering
  \includegraphics[width=\linewidth]{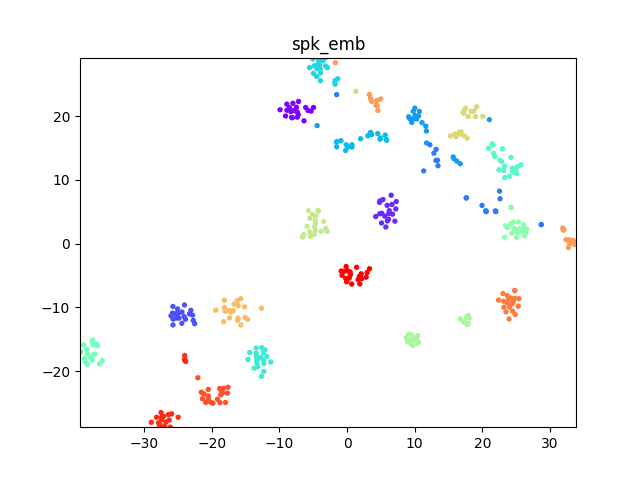}  
  \caption{}
  \label{fig:sub-first}
\end{subfigure}
\begin{subfigure}{.23\textwidth}
  \centering
  \includegraphics[width=\linewidth]{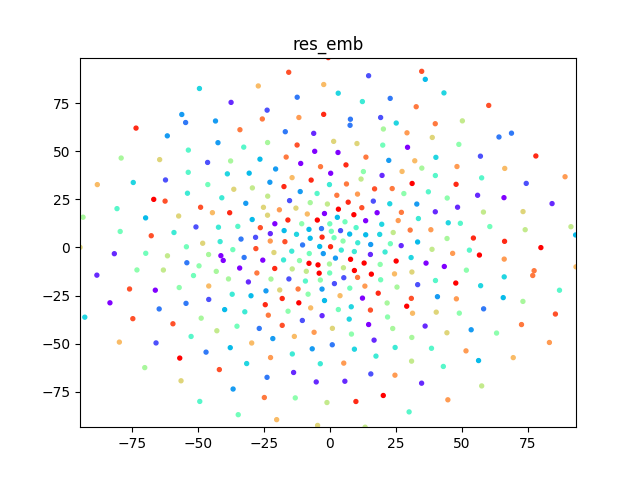}  
  \caption{}
  \label{fig:sub-second}
\end{subfigure}
\newline
\begin{subfigure}{.23\textwidth}
  \centering
  \includegraphics[width=\linewidth]{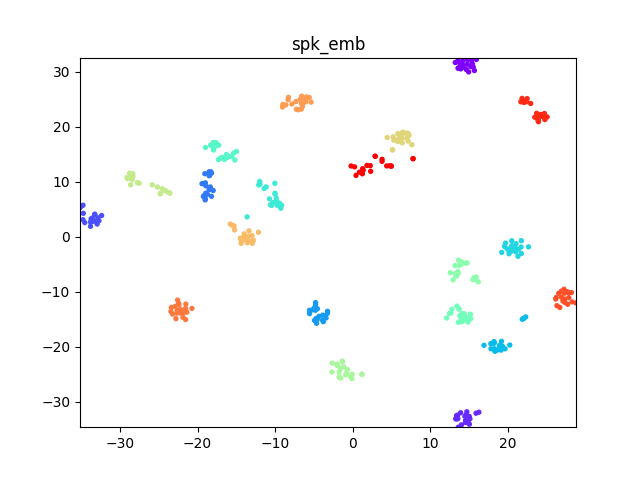}  
  \caption{}
  \label{fig:sub-third}
\end{subfigure}
\begin{subfigure}{.23\textwidth}
  \centering
  \includegraphics[width=\linewidth]{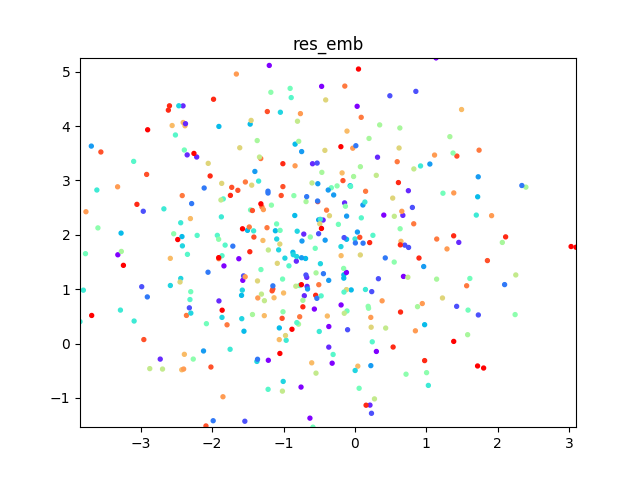}  
  \caption{}
  \label{fig:sub-fourth}
\end{subfigure}
\caption{t-SNE plot of extracted embeddings: extracted from 10 speaker and 20 utterances each and each color corresponds to a different speaker. {\normalfont (a)} and {\normalfont (b)} are extracted from baseline model. {\normalfont (c)} and {\normalfont   (d)} are from our proposed model.}
\vspace{-8pt}
\label{fig:fig}
\end{figure}

Figure~\ref{fig:fig} illustrates t-SNE plots~\cite{maaten2008visualizing} for visualization of the effectiveness of the proposed method more concretely.
As shown in Figure~\ref{fig:sub-first} and Figure~\ref{fig:sub-second}, the proposed model also effectively disentangles speaker-related and speaker-unrelated information.
Moreover, compared to the baseline with proposed model in Figure~\ref{fig:sub-first} and Figure~\ref{fig:sub-third}, our method shows more densely clustered identities with small variance.

Through the results of experiments, we proved that mutual information loss and identity change loss is helpful in learning the clearly disentangled features for speaker recognition.

\section{Conclusion}
In this paper, we present a novel disentanglement training scheme to estimate more informative speaker embedding vectors for robust speaker recognition.
Our method is built upon auto-encoder frameworks with two encoders and trained via mutual information and identity change loss, which extracts more discriminative representations by reducing the variance in the intra-cluster.
Experimental results demonstrated that our algorithm achieved improved EER compared to the baseline method.
Through ablation experiments, we demonstrated the impact of each criterion to the overall performance.

\noindent\textbf{Acknowledgements.} 
This research is sponsored by Naver Corporation.

\pagebreak
\bibliographystyle{IEEEtran}
\bibliography{mybib}

\end{document}